\documentclass[preprint,aps,floats,nofootinbib,amssymb]{revtex4}
\pdfoutput=1 
\usepackage[T1]{fontenc} 
\usepackage{hyperref}
\usepackage[utf8]{inputenc}
\usepackage{amssymb}
\usepackage{latexsym}
\usepackage{graphics}
\usepackage{graphicx}
\usepackage{feynmf}
\usepackage{color}
\usepackage[sort&compress]{natbib}
\usepackage{epsf,epsfig}
\usepackage{amsmath}
\usepackage{hyperref}
\usepackage{subfigure}
\usepackage{mathrsfs}

\newcommand{\be}{\begin{equation}}
\newcommand{\ee}{\end{equation}}
\newcommand{\ba}{\begin{eqnarray}}
\newcommand{\ea}{\end{eqnarray}}

\newcommand{\bw}{\begin{widetext}}
\newcommand{\ew}{\end{widetext}}

\begin{document}         
\title{\vspace*{1.in}\large
Do primordial quark pellets solve the dark matter puzzle?}
\vspace*{0.5cm}

\author{Dom\`enec Espriu\footnote{espriu@icc.ub.edu}}
\affiliation{
Departament de F\'isica Qu\`antica i Astrof\'isica and
Institut de Ci\`encies del Cosmos (ICCUB), \\
Universitat de Barcelona, 
Mart\'i Franqu\`es 1, 08028 Barcelona, Spain}

\begin{abstract}
  We show that primordial quark pellets (PQPs)---ultra-dense quark‑matter mini-stars formed at $T\sim  1$ GeV ---could
  naturally arise in a radiation dominated early universe if rare baryon overdensities are produced by contracting
  Peccei–Quinn domain walls or similar super-horizon structures. Solving the Tolman–Oppenheimer–Volkoff equation
  with a hot quark equation of state, we find stable solutions with a maximum theoretical mass of 10$^{-2}$ $M_\odot$
  and radii of order 100 m, although the formation mechanism favours much lighter objects that would indeed be dominant.
  Once formed, PQPs cool and evolve into either mini neutron stars or stable strange-matter nuggets, depending on the QCD ground state.
  The mechanism we present could suffice to reproduce the present dark‑matter density without altering Big Bang Nucleosynthesis
  or requiring a posterior entropy injection.  PQPs may evade current microlensing constraints if their masses are below 10$^{-8} \ M_\odot$,
  a range that in view of their mass function could easily accommodate a large fraction of dark matter. 
  PQPs thus potentially constitute a conservative, Standard‑Model‑based, and observationally testable candidate.

\end{abstract}

\maketitle

\section{Introduction}
The purpose of this article is to convince the reader that micro-stars may be a viable candidate for
cold dark matter. These micro-stars are composed of extremely dense primordial quark pellets (PQP) frozen out at
a temperature around 1 GeV in the primordial universe, progressively cooling and evolving to become either
mini neutron stars or quark-matter nuggets of the sort proposed several years ago by Witten and other authors\cite{nuggets}.
Their mass would be in the sub-planetary range or even lighter.

There is no need to insist on the fact that the nature of dark matter is one of the greatest puzzles of
cosmology\cite{dark}, with a profound impact on various areas of fundamental physics. To this date no compelling evidence has been
found in favour of any of the scores of proposals put forward.

A recent article by Krnjaic, Rocha and  Xiao\cite{KRX} suggested a novel cosmological scenario in which neutron stars
could plausibly form in the early universe provided that baryogenesis provided a baryon asymmetry much larger than is
conventionally assumed. Taking into account
that the observed baryon asymmetry is yet to be convincingly explained by the Standard Model, it does
not seem too audacious to propose that it could actually be much larger at earlier times. In \cite{KRX} it was proposed that the
baryon density could locally exceed the minimum neutron star mass before big bang nucleosynthesis and store a sizeable amount
of what we would call cold dark matter without disturbing big bang nucleosynthesis (BBN), which is well established\cite{Fields2020}.

For the scenario in \cite{KRX} to be plausible, the formation of the primordial neutron stars should take place before BBN, but not much earlier as
neutrons would be unstable and the temperature too high for star stability, thus requiring a certain amount of fine tuning.
As said, the scheme would need an enormous baryon density that
would dominate the primordial universe. To restore the known
evolution, including BBN, a large injection of entropy would be eventually needed. In \cite{KRX} it is claimed that the resulting primordial neutron
stars could be as light as 0.1 solar masses, but
typically larger, something that could possibly have difficulties with existing limits, particularly if one defends the point of view that
most dark matter should be of this type.

However, the proposal just described is attractive for one reason: one does not need to assume any physics beyond
the standard model. All particles
and interactions are known and the main assumption is that of a very large primordial baryon asymmetry.

We concluded that a modification of the proposal in \cite{KRX} is more natural
and in many senses more satisfactory. One needs to assume a larger primordial baryon density, but not as large as the one
required by \cite{KRX} so that the primordial universe is not overwhelmingly dominated by baryons. As a consequence
the picture becomes more plausible. The resulting objects are mini stars, with a theoretical upper limit around 10$^{-2}$ $M_\odot$ but could be
10$^{-5}$ or 10$^{-7}$ $M_\odot$ or even less. The upper limit fits well with
the horizon constraints at the time of their creation. In addition, no entropy injection mechanism will be required prior to BBN.

Since no fully satisfying mechanism
for baryogenesis exists to this date, the present baryon to photon density has to be assumed by fiat.
Accordingly, we can accept the required initial baryon density in the understanding that
the excess baryons should be sequestered in PQPs and it will not affect the subsequent evolution. The primordial universe
will be radiation dominated. In order to seed the PQPs, we will need sufficiently large inhomogeneities in the baryon density
in the $T=1$ GeV epoch. These could be provided by the coalescence and contraction of domain walls originally due to the
breaking of the Peccei-Quinn symmetry\cite{PQ}. While this is strictly speaking beyond the minimal standard model, it has become
a widely accepted mechanism, likely to be relevant in the early universe.

For reasons that will be clear in a moment, the temperature of formation of PQPs
cannot be larger than the charm quark mass, i.e. $T < 1.27$ GeV. It has to be
above the QCD phase transition at around $T\approx 200$ MeV that guarantees a safe separation with BBN. We have taken $T=1$ GeV,
a temperature that fits all the above conditions and such that everything is calculable. We will also explore the possibility of
the creation of the mini-stars taking place a bit later, at around $T\approx 400$ MeV. This will give us an idea about the
dependence of the results on the initial temperature.

The mini stars so created are extraordinarily dense, much smaller and substantially denser than a typical neutron star, eventually
becoming free streaming objects that have no other interactions but gravity. The question as to how they may have evaded
detection will be discussed too.

This work focuses on the microphysics and cosmological viability of PQPs, and the possible formation mechanisms are dealt
with only qualitatively to motivate the viability of the proposal.
They are discussed in more detail in a companion paper \cite{companion}.

\section{Summary of equations and boundary conditions for primordial quark pellets}

In this study, we investigate the macroscopic structure of a hot baryon star embedded in an early-universe ambient background
characterized by a temperature $T=1$ GeV. The relevant baryons are quarks of the $u, d, s$ flavour. A mixture of these
flavours in approximate equilibrium is neutral by construction. This electric neutrality would be lost if
the charm quark is active making impossible the aggregation of quarks into neutral lumps that, incidentally, would
also carry a large weak charge - just as a neutron star.

At such temperatures, the effective number of degrees of freedom is $g_* = 61.75$, a factor that multiplies the energy
and pressure of the ambient and makes both of them quite large. As to the system formed by the aggregation of hot quarks, whose
stability we will consider in a moment, is governed by a highly relativistic,
degenerate quark equation of state with thermal corrections. 

The internal energy density $\epsilon(n, T)$ and pressure $P(n, T)$ as functions of the baryon number density $n$ and
local temperature $T$ are given by
\begin{equation}
\epsilon(n, T) = \frac{9}{4} \pi^{2/3} n^{4/3} + 3^{1/3} \pi^{4/3} n^{2/3} T^2
\end{equation}
\begin{equation}
P(n, T) = \frac{3}{4} \pi^{2/3} n^{4/3} + \frac{1}{3} \cdot 3^{1/3} \pi^{4/3} n^{2/3} T^2.
\end{equation}
The total quark density will be $3n$ and each flavour number density will be equal to $n$; i.e. $n_u=n_d=n_s=n$.

The alert reader has noticed that no degrees of freedom other than baryons are considered in the interior, whereas photons and leptons
contribute to the exterior ambient along with quarks. This treatment will be justified in Section III.D and also in 
Appendix B.

At the surface boundary $r = R$, hydrostatic equilibrium demands that the internal pressure of the quark proto-star
match the external radiation pressure exerted by the surrounding plasma universe
\begin{equation}
P_{\text{internal}}(n_R, T) = P_{\text{external}}(T) = \frac{g_* \pi^2}{90} T^4
\end{equation}

Substituting the environmental parameters ($T = 1\text{ GeV}$, $g_* = 61.75$), equating the pressures results
in a definite boundary baryon number density:
\begin{equation}
n_R \approx 1.793\text{ GeV}^3 \quad (\approx 233.8\text{ fm}^{-3})
\end{equation}
This exceptionally high boundary density ($\approx 1500$ times normal nuclear saturation density) self-consistently
justifies the highly degenerate, asymptotically free treatment of the quark core, and the thermal $T^4$ corrections
remain safely negligible inside the bulk of the quark proto-star.

At the boundary density $n_R \simeq 1.8\,\mathrm{GeV}^3$, the per-quark chemical potential is 
$\mu \simeq 2.6~\mathrm{GeV}$, where asymptotic freedom ensures a small coupling 
$\alpha_s(\mu)\approx 0.25\text{--}0.35$. The leading perturbative correction to the quark 
pressure,  $P \;\rightarrow\; P\left(1-\frac{2\alpha_s}{\pi}\right)$, modifies the EOS by at most $\sim 15\%$.
This corresponds to a shift well below $10\%$ in the predicted PQP masses. These corrections are subdominant compared to other model 
uncertainties, and we neglect them for clarity.

To prepare the system for general relativistic stellar structure integration via the Tolman-Oppenheimer-Volkoff (TOV)
equation, we define a dimensionless density variable normalized to the surface boundary value
\begin{equation}
x \equiv \frac{n}{n_R}
\end{equation}

Expressing the equation of state (EOS) explicitly in terms of $x$ at $T = 1\text{ GeV}$ sets the value of the parameters.
Units are GeV$^4$
\begin{equation}
\epsilon(x) = 10.53 x^{4/3} + 9.78 x^{2/3}
\end{equation}
\begin{equation}
P(x) = 3.51 x^{4/3} + 3.26 x^{2/3}
\end{equation}

Taking the derivative of pressure with respect to our dimensionless parameter $x$ yields the final analytic mapping
required for integration
\begin{equation}
\frac{dP}{dx} = 4.68 x^{1/3} + 2.17 x^{-1/3}
\end{equation}
The relevant TOV equation is
\begin{equation}
x'= -4G\epsilon(r)(\frac{dP}{dx})^{-1}\frac{m(r)+ 4 \pi r^3 P(r)}{r (r - 2 G m(r))},
\end{equation}
where $P(r)= \epsilon(r)/3$ and
\begin{equation}
  m(r)= \int^r 4 \pi y^2 \epsilon(y) dy
\end{equation}
These equations require some initial conditions $x(r=0)=x_c$, $m(r=0)=0$, which need to be treated with some
care as the equation is singular at $r=0$, as is well known.

\begin{figure}
 \centering
 \includegraphics[scale=0.6]{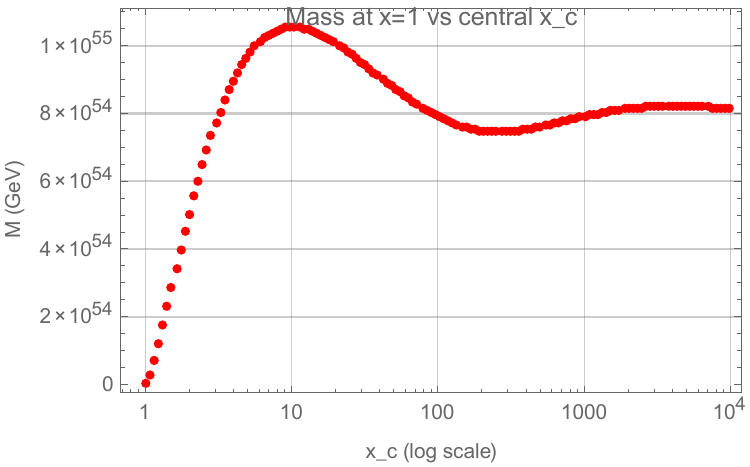}
 \includegraphics[scale=0.6]{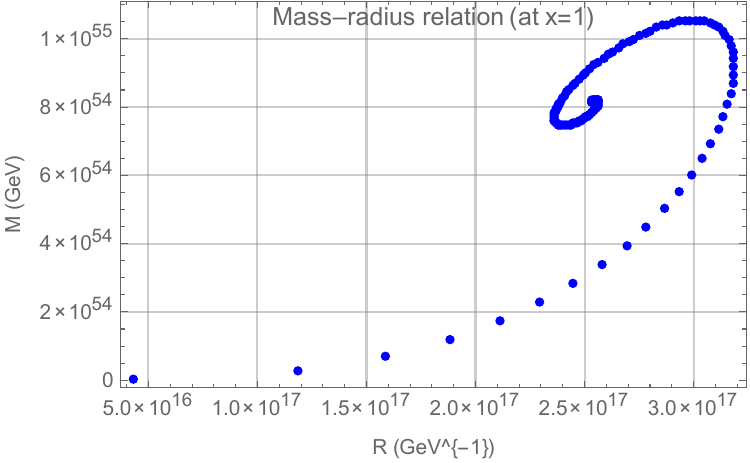}\\
 \caption{Mass versus central density resulting from the integration of the relativistic TOV equation (left).
   Mass versus radius (right). Masses  are in GeV and distances in GeV$^{-1}$}
 \label{fig:TOVresults}
\end{figure}

 Unlike standard cold compact objects in a vacuum where the integration boundary condition terminates at $P(R) = 0$,
  this hot quark star terminates precisely when the internal density drops to the environmental equilibrium threshold, meaning
$x(R) = 1$.
Sampling a series of values for $x_c$ one obtains the results depicted in  figure (\ref{fig:TOVresults}).
From these results we conclude that the maximum mass is 0.013 $M_\odot$ and the maximum
radius is 88 meters (the maximum of the mass does not correspond to the largest radius). The maximum central density allowed
is a little over 7 times the density at the boundary.
The Schwarzchild radius of this object is 38 meters for the largest mass.

We emphasize that there is very little room to modify these results. Once the temperature is fixed, the relativistic equations
do the rest. As announced the maximum mass lies well below the range advocated in \cite{KRX}.  For a comparison we quote
the analogous results for $T=400$ MeV:  maximum mass= 0.082 $M_\odot$, maximum radius= 520 m. The increase with decreasing temperature
reflects the fact that the maximum mass configuration at lower $T$ has
a much larger baryon number. Yet, a pellet of fixed baryon number becomes slightly lighter as the universe cools down (see Appendix A).

We also studied the solutions to the TOV equation at $T=200$ MeV as they would allow us to follow the evolution of PQPs created
earlier (see Appendix A). Since we would be right at the QCD phase transition region, the number of effective degrees of freedom
contributing to the radiation background has been reduced to photons, electrons, positrons and neutrinos which gives
$g_*= 10.75$. Being fully in the hadronic phase, the solutions to the TOV equation are very familiar: they have a maximum mass
of 0.7 $M_\odot$ and 4 to 5 Km in size, as befits a neutron star when only the degeneracy pressure contributes. However, this maximal
PQP really is of no interest to us; what matters is the low-mass tail of the 200 MeV TOV curve, which describes the endpoint of the pellets formed at 1 GeV. 

All solutions for $x_c$ below the value when $dM_{PQP}/dx_c=0$ are known to be
stable and physically acceptable solutions. For reasons that will be
apparent later we also plot the region of parameters corresponding to low
values of the central density parameter $x_c$, which yield the lightest objects.
\begin{figure}
 \centering
 \includegraphics[scale=0.7]{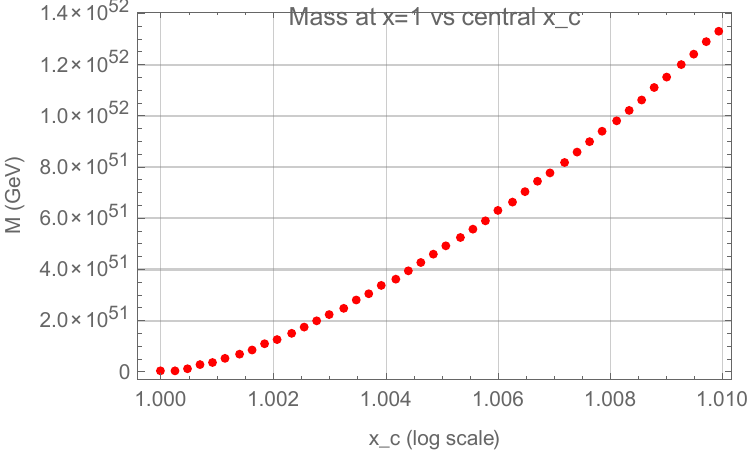}\\
 \caption{Mass versus central density resulting from the integration of the relativistic TOV equation in the region of small masses. The plot depicts PQP in the 10$^{-7}\ M_\odot$ to 10$^{-5}\ M_\odot$ range. Typical radii are up to 10 meters.}
 \label{fig:TOVresults2}
\end{figure}

\section{Cosmological considerations and evolutionary fate}

To assess the physical viability of the hot quark star configurations obtained from the TOV integration
at an ambient temperature of $T = 1\text{ GeV}$, several considerations are in order.

\subsection{Causal Horizon Constraints}
In a radiation-dominated universe, the expansion rate is governed by the Friedmann equation
\begin{equation}
H^2 = \frac{8\pi G}{3}\epsilon_{\text{r}}
\end{equation}
where $G = 6.7083 \times 10^{-39}\text{ GeV}^{-2}$ and $\epsilon_r$ is the ambient energy density
that has been used above to determine the boundary conditions of the mini-star. Assuming again
a standard relativistic plasma with $g_* = 61.75$
degrees of freedom at $T = 1\text{ GeV}$, the ambient energy density is given by
\begin{equation}
\epsilon_{\text{r}} = \frac{g_* \pi^2}{30} T^4 \approx 20.31\text{ GeV}^4
\end{equation}
This yields a Hubble parameter of $H \approx 1.07 \times 10^{-18}\text{ GeV}$. The physical radius of the causal
cosmic horizon ($R_{\text{hor}}$) and the total mass enclosed within the horizon volume ($M_{\text{hor}}$) are
\begin{align}
R_{\text{hor}} &= \frac{1}{H} \approx 9.35 \times 10^{17}\text{ GeV}^{-1} \quad (\approx 185\text{ m}) \\
M_{\text{hor}} &= \frac{4}{3}\pi R_{\text{hor}}^3 \epsilon_{\text{r}} \approx 6.71 \times 10^{55}\text{ GeV} \quad (\approx 0.062\ M_\odot).
\end{align}
To obtain the mass a simple euclidean calculation has been made.
Comparing these bounds to the peak stable configurations found numerically
($M \approx 10^{-2}\ M_\odot$, $R \approx 80\text{ m}$), we find that the primordial quark pellets fit comfortably
within a single causally connected patch ($R < R_{\text{hor}}$) but would consume a hefty $\sim 20\%$ of the locally
available horizon mass.

This also holds in the $T=400$ MeV case because the temperature is lower and the horizon and enclosed energy larger.
Now the horizon radius is about 1150 m and encloses about 0.4 $M_\odot$.

However, if evenly distributed in the average horizon patch, there is {\em not enough} baryon matter. We will discuss this important point
later.

\subsection{Late-time evolution}
As the background universe expands and cools, the primordial quark pellets should decouple from the Hubble flow due to their
internal energy, approximately maintaining a fixed physical scale while surrounding space expands. 
Thermodynamically, the droplets cool via neutrino and photon emission. The total internal pressure profile
progressively loses its second term
\begin{equation}
P(n, T) = 3.51 x^{4/3} + 3.26 x^{2/3} T^2.
\end{equation}
Concurrently, the external boundary pressure vanishes ($P_{\text{ext}} \to 0$), shifting the surface boundary
condition from $x(R)=1$ to the vacuum limit $x(R) \to 0$. To compensate for the loss of thermal pressure support,
the configuration adjusts quasi-statically. For a fixed baryon number the mass decreases by a
factor of $\sim 4 $ and the radius grows by a factor of $\sim 8$ between formation and the QCD
transition (see Appendix A). The maximal configurations behave differently, contracting slightly,
but they are marginal.  Below 200 MeV the PQPs should be frozen. 

The final stage, however, cannot be determined by our arguments.
Depending on the true ground state of QCD at zero temperature, two evolutionary paths exist:
\begin{enumerate}
\item \textbf{Hadronization:} The quarks confine into nucleons, causing the core to transition into standard nuclear matter,
  yielding a population of ultra-compact primordial neutron stars but much lighter than the ones advocated
  in \cite{KRX}
\item \textbf{Witten's strange matter hypothesis:} If three-flavor strange quark matter is the absolute ground state of
  the strong interaction, the mini-star
  freezes permanently as a stable, zero-temperature \textit{strange quark mini-star}\cite{nuggets}.
\end{enumerate}
Both possibilities are equivalent from an astronomical point of view. We emphasize here again that
the masses and radius quoted in the previous section are the maximum ones allowed by general relativity
and the ambient constraint. Typically
one should expect much smaller masses.

\subsection{Abundance and dark matter accounting}
Because these primordial quark pellets are bounded, their average energy density dilutes as matter
($\epsilon_{DM} \propto a^{-3}$), whereas the background plasma dilutes as radiation
($\epsilon_r \propto a^{-4}$).
Consequently, they rapidly grow to dominate the energy density of the universe. 

The current dark matter/radiation density ratio is  $\epsilon_{DM} / \epsilon_r \approx 2880$ and
extrapolating this backward to $T = 1\text{ GeV}$
dictates that dark matter must comprise a mere fraction of $\sim 10^{-10}$ of the total cosmic energy density at that epoch.
Let us see how this figure comes about.

If $a$ is the cosmological factor, we know that dark matter decreases like $a^{-3}$ and radiation like
$a^{-4}$ and that $a$  goes like $1/T$ because the expansion is adiabatic. Then a simple calculation
gives
\begin{equation}
\frac{\epsilon_{DM}}{\epsilon_r}(T)= \frac{\epsilon_{DM}}{\epsilon_r}(T_0) \times \frac{T_0}{T}  
\end{equation}
This gives $6.8 \times 10^{-10}$. Taking into account the change in degrees of freedom reduces the figure
to $5.8 \times 10^{-10}$.

The required dark matter mass per horizon volume is roughly
\begin{equation}
M_{\text{DM, hor}} = 5.8 \times 10^{-10} \times M_{\text{hor}} \approx 3.6 \times 10^{-11}\ M_\odot
\end{equation}
Given a maximal mass of $M_{\text{PQP}} \approx 10^{-2}\ M_\odot$, the required formation abundance to account
for $100\%$ of the observed cold dark matter would be
\begin{equation}
\frac{M_{\text{DM, hor}}}{M_{\text{PQP}}} \approx  10^{-9} 
\end{equation}
primordial quark pellets per horizon
This indicates that the model is  efficient. If correct, it would require only one of the largest primordial
quark pellet of 10$^{-2}$ $M_\odot$ per $10^9$ horizon volumes; to produce a 10$^{-8}\ M_\odot$ pellet one would need some
10$^3$ horizons, or in fact any combinations that add up to the required concentration of baryonic matter
at $T=1\text{ GeV}$ to fully resolve the modern dark matter problem, leaving the homogeneous background FLRW evolution unperturbed,
including BBN, if all the excess baryons are sequestered in PQPs.

However, the average horizon volume does not contain sufficient baryons, as is obvious from the previous calculation. This
implies that some horizons at 1 GeV have to be much richer in baryons than the average. We will discuss later how this could come about.

To be complete, let us check some of those numbers if the lower temperature $T\approx 400$ MeV is selected.
For this lower temperature the number of degrees of freedom is still the same $g_*(T) = 61.75$. 
The evolution of the present density ratio to that epoch is exactly the same as before and
at $T = 0.4\text{ GeV}$, the total radiation density is now
\begin{equation}
\epsilon_{\text{r}}(T=400 \ \text{MeV}) = \frac{61.75 \cdot \pi^2}{30} (0.4)^4 \approx 0.520\text{ GeV}^4
\end{equation}
The total mass enclosed within this causal sphere is
\begin{equation}
M_{\text{hor}} = \frac{4}{3}\pi R_{\text{hor}}^3 \epsilon_{\text{r}} \approx 4.36 \times 10^{56} \text{ GeV} \approx 0.4\ M_\odot
\end{equation}
Multiplying the horizon mass by the dark matter energy fraction at 400 MeV yields the exact required dark matter
allocation per horizon volume
\begin{equation}
M_{\text{DM, hor}} = (1.5 \times 10^{-9}) \times 0.4\ M_\odot = 5.8 \times 10^{-10}\ M_\odot
\end{equation}
Assuming a typical stable droplet near the upper limit of $M_{\text{PQP}} \approx 10^{-2}\ M_\odot$, the required
formation frequency across the early universe is
\begin{equation}
  \text{Abundance} = \frac{M_{\text{DM, hor}}}{M_{\text{star}}} \approx 6 \times 10^{-8} \text{ PQP per horizon}
\end{equation}
Inverting this probability reveals that one large PQP  must nucleate in  $10^7$ causal horizons at $T= 400$ MeV
to account for $100\% $ of the observed cold dark matter today, a figure that may
seem quite reasonable too,
but of course the same caveat as before applies. The viability of the model requires a larger baryon abundance in some
places to seed the formation of the mini-stars.
Below we suggest a plausible mechanism.

\subsection{A sweep-and-collect mechanism for baryon concentration}

It is far beyond the scope of this work to describe baryogenesis. We simply assume that a sufficiently large baryon charge is produced
early enough. 

The local concentration of baryons required to seed, for instance, a stable $10^{-8}\ M_\odot$ pellet within a single $1\text{ GeV}$
horizon (where the average baryon content is $\sim 10^{-11}\ M_\odot$) cannot be
explained by statistical fluctuations. Rather, it could find 
a natural explanation if the Peccei-Quinn phase transition\cite{PQ,other} occurs at a much higher scale.

After the Peccei-Quinn transition occurs, $\theta$ is uncorrelated between distant horizons.
Domain walls separate different $\theta$ vacua after the breaking of the Peccei-Quinn symmetry. If the scale of the
transition is large, domain walls have an enormous surface tension that partially counteracts the cosmological flow and
will eventually force them to coalesce (and eventually decay into axions).

We propose that the required baryon overdensities relative to the horizon average at
$ T= 1 $ GeV can be achieved through the contraction of super-horizon bubbles or domain walls. If a closed bubble with physical
radius $ R_i $ forms at an earlier epoch with temperature $T_i$ as it re-enters the horizon
sweeps up and concentrates ambient baryons into a dense central region. For instance, a bubble with linear size
$ \sim 10^6 $ horizon volumes at the earlier epoch, the enclosed mass is in principle sufficient to seed a PQP of mass
$ 10^{-5} M_\odot $. This mechanism is analogous to the domain wall collapse scenario for primordial black hole
formation \cite{Liu2019}, but with the ``collapsed'' object remaining sub-Chandrasekhar and stabilizing as a PQP.

Causality indicates that the supra-horizon correlations must have formed
around the inflationary period.  An alternative (but without a definite physical picture) could be
that the primordial overdensities arise from scale-invariant inflationary perturbations.

If PQPs form via the sweep-and-collect mechanism of a brownian domain wall network in the scaling regime, the number
density of collapsing bubbles of radius $R$ scales as $dN/dR \propto R^{-4}$ \cite{shellard}. Since the collected
baryon mass is $ M_{DM} \propto R^3 $, this implies a mass function
\begin{equation}
\frac{dN}{dM_{PQP}} \propto M^{-2}_{PQP}
\end{equation}
This power-law spectrum is skewed toward low masses. 
Once a minimum PQP mass can be established, it would be possible to determine the proportionality constant from the
measured dark matter density
and that would allow us to model the size of the initial bubbles and connect with the Peccei-Quinn transition. 

We mention domain‑walls as a plausible seeding mechanism for the rare baryon overdensities required by PQP formation;
a quantitative treatment of sweep‑and‑collect (bubble statistics, sweep efficiency, and diffusion timescales) is beyond the scope
of this paper and is addressed in the companion study \cite{companion}.

 If instead the primordial overdensities arise from scale-invariant inflationary perturbations,
 the Press-Schechter form\cite{PS} gives $ dN/dM_{PQP} \propto M^{-1.66}_{PQP} $, which still populates the low-mass tail
 but with slightly more weight at higher masses.

 The sweep-and-collect mechanism justifies why only baryons are considered in the PQP interior, at least ideally.
 The confining domain wall is a configuration of the axion field. The
axion couples to strongly interacting matter through the QCD axial anomaly
and to photons through the electromagnetic anomaly
(suppressed by $\alpha_{\rm EM}$); it does not couple to leptons at tree level.
As a result, the wall is opaque
to quarks but transparent to photons, and leptons. The pellet is
therefore an \emph{open} system with respect to those species.

However, as the domain walls start collecting quarks, a growing part
of the internal energy and pressure is due to them, implying that
the fractional pressure due to photons and leptons decreases. In the limit
this would become very small. In other words, exactly in the
same way that the baryon chemical potential is discontinuous
across the surface of the PQP, so is the lepton chemical potential but
in the opposite sense. Outside the pellet there is perfect thermal equilibrium
of all degrees of freedom.

The role of thermal gluons is more delicate and it is discussed in Appendix B.

\section{Observability of the PQP mini-stars}

Primordial quark pellets formed  at the $T \sim 1\text{ GeV}$ scale apparently represent a compelling macroscopic dark matter candidate.
If the formation efficiency yields typical pellet masses
below  $M_{\text{PQP}} \sim 10^{-8}\,M_\odot$  these objects may escape many standard baryonic constraints while avoiding
immediate collapse into primordial black holes. 

One should ask how abundant would they be. The local dark matter density is about 0.011 $M_\odot$ per cubic parsec. If one
sticks to the absolute upper limit of 10$^{-2}$ solar masses, this yields about 1 PQP per cubic parsec,
which is slightly above the density of stars
in our local environment (about 0.2 per cubic parsec). If the typical mass is $10^{-5} M_\odot$ we would be talking about thousands
of PQP per cubic parsec. Even though they are relatively numerous, as one can infer from this discussion, they are small so that
the total dark matter density is actually lower than normal baryonic mass in the galaxy plane. However, since they are free
streaming and do not interact, they add up to the expected
dark mass density when the galaxy halo is considered. However, the expected distribution does definitely depend on the dark matter modelization.

The primary direct observational probe for macroscopic dark matter objects in the sub-solar mass range is microlensing. 
Optical microlensing surveys of the Andromeda galaxy (M31) by the Hyper Suprime-Cam (HSC)
  and the galactic bulge by OGLE placed stringent limits on the fraction of dark matter $f \equiv \epsilon_{\text{PQP}}/\epsilon_{\text{DM}}$ locked
  in compact objects \cite{Niikura2019, Niikura2019ogle}.
  Traditional baryonic MACHO\cite{machos} halos (with characteristic masses between 0.1 $M_\odot$ and 1 $M_\odot$)
  were excluded  by the multi-year null results of the EROS-2 survey \cite{Tisserand2007} ruling out objects larger than $10^{-7} \ M_\odot$
  as the dominant component of dark matter.

  On the other end of the spectrum, Witten's nuggets with masses ranging from $ 10^{-25}\ M_\odot$ to $10^{-16} \ M_\odot$ were
  all but ruled out once it was established that the confinement transition was not of first order. In addition, fast evaporation rates would rule them out too.

In between these two extremes, the compact configuration of PQPs may allow them to escape undetected. It is known that
in the sub-planet and Earth-mass regimes ($M < 10^{-3}\ M_\odot$), the angular Einstein radius falls drastically below the
angular radius of monitored giant source stars in the galactic bulge. Consequently, the severe finite-source size damping
outlined in \cite{Witt1994,Gould1997} washes out the geometric magnification curve entirely, rendering the population optically
invisible.

Nevertheless, recent results \cite{OGLEnew} show that the limit could actually be fairly stringent: compact
objects with $ M > 1.4 \times 10^{-8}\ M_\odot$ could comprise 1\% of dark matter at most. However,
if the formation  mechanism shifts the average pellet mass below that limit, they fall below the
finite-source size effect threshold of optical surveys, opening up an unconstrained parameter space window where
PQPs can potentially constitute a large fraction of the dark matter. It should also be mentioned that the actual OGLE limit depends
to a non-negligible extent on the model chosen for the distribution of dark matter.

If the mechanism producing the required baryon inhomogeneities takes place around  $1\text{ TeV}$ and is a strongly
first-order phase transition, it would inevitably generate a stochastic
gravitational wave background. The characteristic peak frequency scales directly with temperature:
\begin{equation}
\omega_{\text{peak}} \approx 1.6 \times 10^{-2} \text{ Hz} \left(\frac{T_{*}}{1\text{ TeV}}\right) \left(\frac{g_{*}}{100}\right)^{1/6}
\end{equation}
For $T \sim 1\text{ TeV}$, this places the signal precisely in the mHz to Hz band, making it a prime target for next-generation
space-based interferometer LISA. \cite{Caprini2016,Caprini2020}.

\section{Conclusions}

The proposal suggested in \cite{KRX} achieves primordial neutron star formation by assuming a vastly larger
baryon asymmetry at early times ( $\eta \gg 10^{-10}$ ), such that baryons dominate the universe and gravitational collapse
is ubiquitous. This requires a subsequent entropy injection event to restore the observed $\eta \sim 10^{-10} $ before BBN.
In contrast, our PQP scenario operates within the standard cosmological framework, radiation-dominated at $T \sim 1 $ GeV.
We do not require a large initial asymmetry or a post-formation entropy injection. Instead, we rely on a local concentration mechanism
(sweep-and-collect) to gather the necessary baryons into rare, compact objects. While our formation mechanism
requires new physics to some extent (e.g., domain walls or inflationary bubbles), it avoids the fine-tuning associated with entropy
injection, neutron star stability issues,  and leaves BBN intact.

In summary, we have shown that primordial quark pellets—ultra-dense, sub-planetary-mass or less objects formed at $ T \sim 1 $ GeV
from the collapse of super-horizon bubbles— could represent a viable cold dark matter candidate. They require no new physics
beyond a mechanism for local baryon concentration, operate within the
standard baryon asymmetry, and may evade existing microlensing constraints
if the mass function is steep enough. Future gravitational wave detectors like LISA might directly probe the formation epoch,
and improved microlensing surveys in the sub-Earth-mass regime will test the low-mass tail of the predicted spectrum.

\section{Acknowledgements}
 The financial support from
the State Agency for Research of the Spanish Ministry of Science, Innovation and Universities through the “Unit of Excellence Maria de Maeztu 2025-2028”
award to the Institute of Cosmos Sciences (CEX2024-001451-M) and through project PID2022-136224NB-C21 is acknowledged. We also acknowledge
the support of the Catalan Government through grant 2021-SGR-249. The author thanks P. Mróz for bringing the
very recent OGLE limits to his attention. I thank J. Salvad\'o for early conversations on the subject.

\appendix

\section{Evolution of the primordial quark pellets}
\label{app:evolution}

The late-time evolution of the PQPs discussed in Sec III.B is not immediately obvious.
Do they really survive to the present epoch as Witten's nugget-like objects or
as small neutron stars?
In this appendix we collect some technical details of the late-time evolution
summarized in Sec.~III.B that seem to favour the first possibility.

\subsection{Relaxation time versus Hubble time}

The first fact that needs to be taken into account is that
PQPs are in thermal equilibrium with the environment and this justifies
the quasi-static approximation used throughout Sec.~III.B. The evolution
can be determined by a series of snapshots at various cosmic temperatures
but we do not have to worry about time dependent solutions.

For this to be true the hydrodynamic relaxation time needs to be shorter than the Hubble
time. For a relativistic fluid with sound speed $c_s = 1/\sqrt{3}$,
\be
t_{\rm relax} \simeq \frac{R}{c_s} = \sqrt{3}\,R,
\qquad
t_H = \frac{1}{H} \simeq \frac{M_{\rm Pl}}{1.66\sqrt{g_*}\,T^2}.
\ee
At $T = 1$~GeV, a $10^{-7}\,M_\odot$ pellet has $R \simeq 10$~m and
$t_{\rm relax}/t_H \simeq 0.1$; a $10^{-11}\,M_\odot$ pellet has
$R \simeq 1$~m and $t_{\rm relax}/t_H \simeq 10^{-2}$. Only the
maximum-mass configuration, with $R \simeq 90$~m, has
$t_{\rm relax} \sim t_H$ while the smaller ones (the vast majority) cool
faster.

The fact that there is thermal equilibrium does not imply that the
pellets physically expand with the cosmological flow, and this will be
seen in detail below. Although $R$
grows by a factor of a few during this period, the ratio $t_{\rm relax}/t_H$ still decreases,
so the quasi-static approximation becomes more accurate at later times. 

\subsection{Virial theorem}

For a static, spherically symmetric fluid with external pressure
$P_{\rm ext}$ at the surface, hydrostatic equilibrium implies the virial
relation
\be
3\!\int P\,dV + W = 3P_{\rm ext}V,
\label{eq:virial}
\ee
where $W<0$ is the gravitational potential energy. This follows directly
from integrating $dP/dr = -G\epsilon m/r^2$ against $4\pi r^3$, with the
identification of the resulting boundary term with $3P_{\rm ext}V$ and of
the bulk integral with $W$.

While for $T=1$ GeV one can use right away the equation of state $P= \epsilon/3$,
at lower temperatures we use the MIT bag EOS; defining
$U = \int \epsilon\,dV$, Eq.~(\ref{eq:virial}) becomes
\be
U = |W| + 4BV + 3P_{\rm ext}V.
\label{eq:virial-MIT}
\ee
The internal energy $U$ compensates the gravitational binding
$|W|$, the bag energy $4BV$ (with $B^{1/4} = 200$~MeV), and the work $3P_{\rm ext}V$ done against the
external pressure. Two regimes may follow:
\begin{itemize}
\item \emph{Gravity-bound} ($|W| \gtrsim 3P_{\rm ext}V$): the
configuration is held together by gravity, and as $P_{\rm ext}\to 0$ it
contracts slightly until supported by cold degeneracy pressure. At
$T = 1$~GeV the crossover occurs at
$M_{\rm crit} \sim 10^{-2}\,M_\odot$; that is, at the maximum possible mass,
so this regime is marginal.
\item \emph{Pressure-confined} ($|W| \ll 3P_{\rm ext}V$): the density is
pinned at $n_R(T)$, and at fixed baryon number $N_B$ the radius grows as
$R \propto n_R^{-1/3}$ as the universe cools.
\end{itemize}
At $T = 1$~GeV the bag term is negligible ($B \ll P_{\rm ext}$); at
$400$~MeV and below it is comparable to or dominant over $P_{\rm ext}$.
The bulk of the dark-matter population lies firmly in the pressure-confined
regime; only the rare maximum-mass configurations are gravity-bound.

\subsection{Constant baryon number}

The evolution of each pellet proceeds at constant baryon number $N_B$.
This is not an assumption but a consequence of baryon-number conservation
in the Standard Model. Below the electroweak scale the sphaleron rate is
Boltzmann-suppressed, $\Gamma_{\rm sph} \propto e^{-2\pi/\alpha_W}$, and
no other baryon-number-violating processes are relevant at
$T \sim 1$~GeV. The pellet is effectively a closed system once formed
as far as baryon number is concerned:
surface evaporation is negligible for $N_B \gtrsim 10^{44}$
\cite{nuggetsFJ}, while our pellets have
$N_B \sim 10^{46}$--$10^{55}$ \cite{companion}; accretion from the surrounding plasma
changes $N_B$ by a fractional amount of order $10^{-11}$ over a Hubble
time; and mergers are negligible given the average
spacing. The pellet cools down via photon and neutrino emission;
and weak decays conserve baryon number. The
fixed-$N_B$ trajectory is therefore the correct physical description of
the late-time evolution.

We have considered three temperatures: 1 GeV, 400 MeV (both already discussed
previously) and also 200 MeV.  
The corresponding boundary conditions are
$P_{\rm ext} = g_*\pi^2 T^4/90$, $g_* = 61.75$ at $T = 1$~GeV and
$400$~MeV, and $g_* = 10.75$ (hadronic phase) at $T = 200$~MeV. The
resulting surface densities are
$n_R \simeq 1.79$, $0.116$, and $4.33\times 10^{-3}$~GeV$^3$ at
$T = 1$~GeV, $400$~MeV, and $200$~MeV respectively. As already said, the MIT
bag model EOS is used for the two lower temperatures.

\subsection{Fixed-baryon-number trajectories}

At formation, each pellet is characterized by its baryon number
\be
N_B =  4\pi\,n_R\int_0^R r^2 x(r)\,dr,
\ee
which is conserved during the subsequent evolution. To follow a pellet
from $T = 1$~GeV to $T = 200$~MeV, we solve the TOV equation at each
temperature for a range of central densities $x_c$, compute $N_B$ for
each solution, and match solutions at different temperatures by their
common $N_B$.

\begin{figure}[h]
 \centering
 \includegraphics[width=0.7\textwidth]{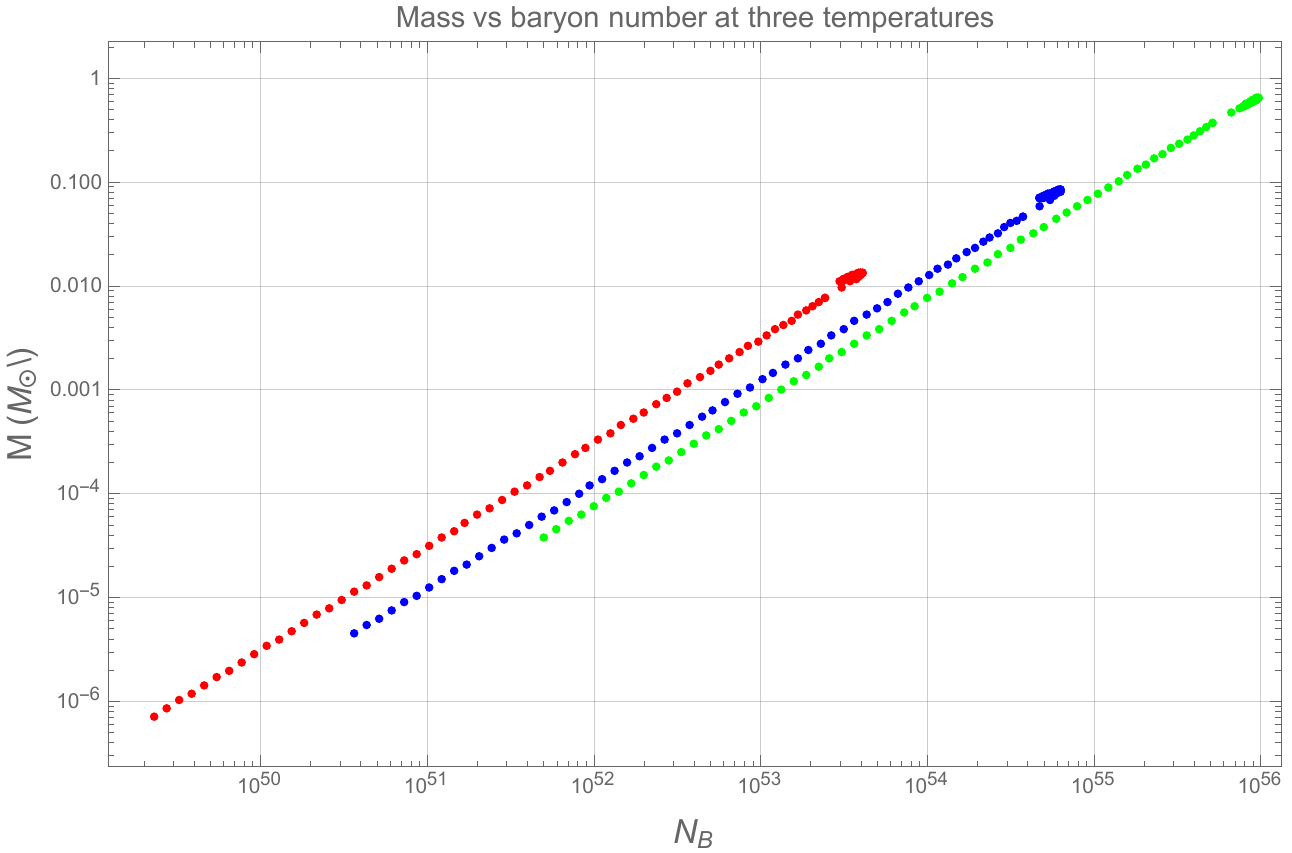}
 \caption{Mass versus baryon number at $T = 1$~GeV (red), $400$~MeV (blue), and
 $200$~MeV (green). The straight lines reflect the pressure-confined regime
 $M \propto N_B$; the bending at the right is the gravitational binding
 near the maximum mass.}
 \label{fig:MvsNB}
\end{figure}

Figure~\ref{fig:MvsNB} shows the mass as a function of baryon number at
the three temperatures. The curves are straight lines in the
log--log plane over most of the range, confirming the pressure-confined
scaling $M \propto N_B$: the energy per baryon is constant along each
curve. The slight bending at the rightmost end marks the onset
of gravitational binding near the maximum mass.

\begin{figure}[h]
 \centering
 \includegraphics[width=0.7\textwidth]{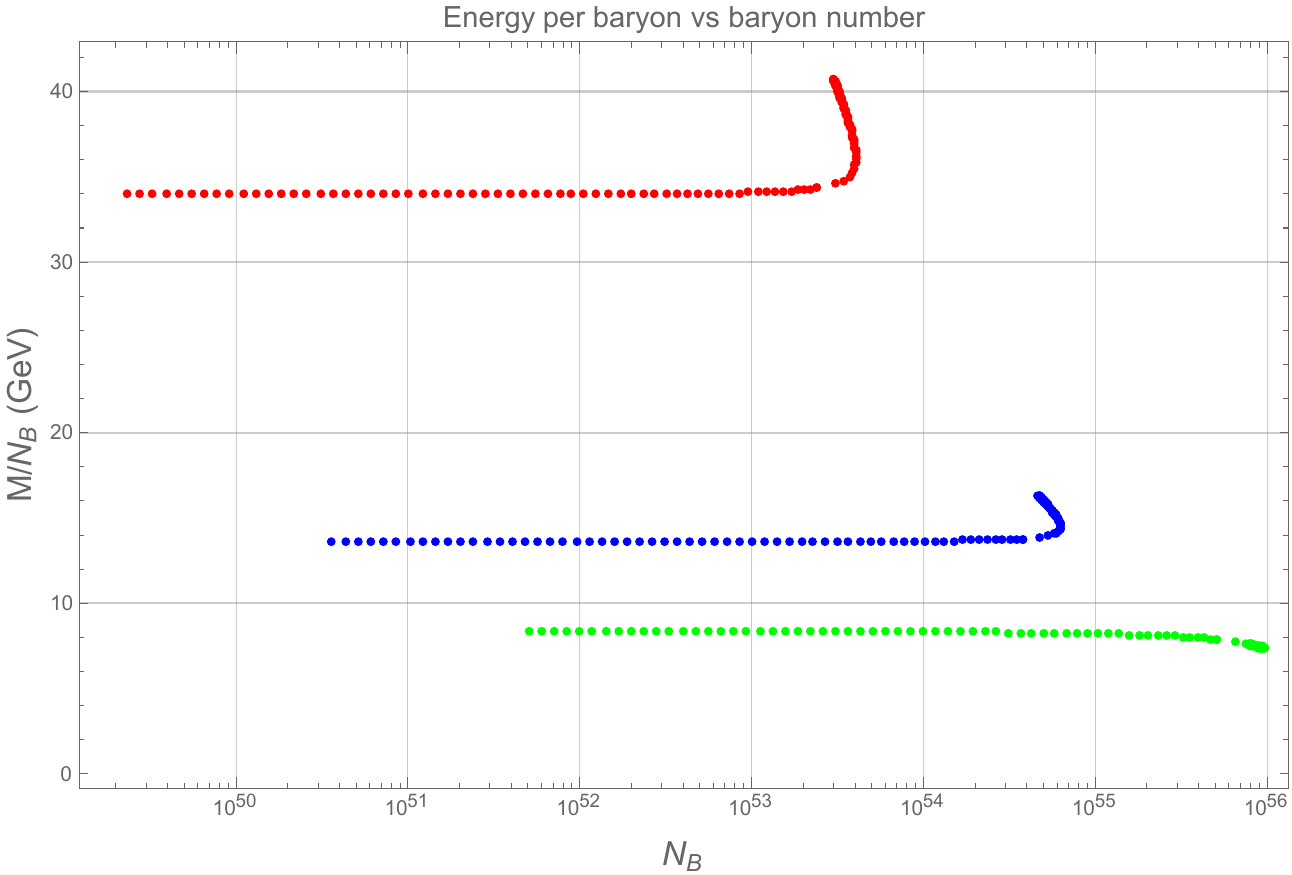}
 \caption{Energy per baryon $M/N_B$ versus baryon number. The horizontal
 plateaus are the pressure-confined regime; the downward bending at
 large $N_B$ is gravitational binding. Color coding as in the previous figure}
 \label{fig:ENB}
\end{figure}

Figure~\ref{fig:ENB} shows the energy per baryon $M/N_B$ as a function of
$N_B$. In the pressure-confined regime (valid almost everywhere) the curves are horizontal, at
levels set by the equation of state.

The study shows that at fixed $N_B$ the mass decreases by a factor of $\sim 4$ between
formation and the QCD transition, while the radius grows by a factor of
$\sim 8$, indeed showing that the compact structures do not follow at all
the cosmic expansion. The PQPs become lighter (thermal energy is
radiated away by photons and neutrinos)
and moderately larger (the external pressure that compressed it disappears). This
is exactly the behaviour expected for a pressure-confined precursor that
freezes into a self-bound strangelet once the QCD transition takes over.

There is a factor of $\sim 50$ between the maximum masses at $1$~GeV and
$200$~MeV but this is not a contraction or expansion effect: it reflects
the fact that the two maximum-mass configurations have different baryon
numbers ($N_B \sim 1.2 \times 10^{54}$ versus
$N_B \sim 5 \times 10^{56}$), because the lower external pressure at
$200$~MeV allows much heavier seeds to remain in equilibrium.

A caveat: at $200$~MeV the surface Fermi energy is
$E_F \sim n_R^{1/3} \simeq 0.16$~GeV, comparable to $T = 0.2$~GeV. The
finite-temperature expansion used throughout this work assumes
$T \ll E_F$, which is only marginally satisfied at this temperature. The
$200$~MeV results are therefore indicative rather than definitive.

\section{Boundary condition and permeability}
\label{app:permeability}

The domain wall is a configuration of the axion field separating two $\theta$ vacua. The axion couples
to gluons through the QCD axial anomaly and to quarks derivatively, but it does not couple to leptons
at tree level and couples to photons only through a loop-suppressed electromagnetic anomaly. As a result,
the wall is opaque to quarks but transparent to photons and leptons.

In the QCD lagrangian the primary coupling of axions is via gluons, but one can make a local chiral
rotation ---a field redefinition--- that removes that term and couples only to quarks. This is quite useful
as we are interested in keeping track of the baryon number and its accumulation inside the PQP. The coupling of
gluons at 1 GeV can be treated perturbatively to a large extent and gives a small modification
to the thermodynamics.

In this basis, gluons do not interact directly with the axion wall and, accordingly, at leading order
can permeate the domain wall just as photons and leptons. As the sweep-and-collect mechanism increases
dramatically the baryon fractional pressure, the contribution from leptons, photons, and gluons too
diminishes. In the limit of an efficient sweep, the interior pressure is dominated by the degenerate quark gas.

However, for the sake of completeness and considering that the gluon coupling is still sizeable we
have contemplated the possibility that the interior of the domain wall
has a gluon density similar to that of the environment at 1 GeV.  

Removing the gluon contribution $g_{\rm gluons} = 8 \times 2 = 16$ from the effective number of relativistic
degrees of freedom at $T = 1$~GeV,
\begin{equation}
g_*^{\rm eff} = 61.75 - 16 = 45.75,
\end{equation}
the effective external pressure is
\begin{equation}
P_{\rm ext}^{\rm eff}(1\ \text{GeV}) = \frac{45.75\,\pi^2}{90} \simeq 5.017\ \text{GeV}^4,
\label{eq:pext-eff}
\end{equation}
to be compared with the value $P_{\rm ext} = 6.772$~GeV$^4$ used in the main text.

The boundary condition is now
\begin{equation}
\frac{3}{4}\pi^{2/3} n_R^{4/3} + \frac{1}{3}\cdot 3^{1/3}\pi^{4/3} n_R^{2/3} T^2 = P_{\rm ext}^{\rm eff}.
\end{equation}
At $T = 1$~GeV, with $u \equiv n_R^{2/3}$ this becomes
\begin{equation}
1.60875\,u^2 + 2.21201\,u - 5.017 = 0,
\end{equation}
whose positive root is $u = 1.2076$, hence $n_R \simeq 1.327\ \text{GeV}^3 \simeq 173\ \text{fm}^{-3}$,
This is $26\%$ lower than the value $n_R = 1.793$~GeV$^3$ quoted in the main text.
The surface density remains enormously high --- about $10^3$ times nuclear saturation --- so the degenerate, asymptotically
free treatment of the quark core remains justified. The quark chemical potential at the surface is
\begin{equation}
\mu_f = (\pi^2 n_R)^{1/3} \simeq 2.36\ \text{GeV},
\end{equation}
and $\alpha_s(\mu_f) \simeq 0.29$ remains in the perturbative regime.

Expressing the EOS in terms of $x = n/n_R$ at $T = 1$~GeV, with units GeV$^4$,
\begin{align}
\epsilon(x) &= 7.04\,x^{4/3} + 8.02\,x^{2/3}, \\
P(x)       &= 2.35\,x^{4/3} + 2.67\,x^{2/3}, \\
\frac{dP}{dx} &= 3.13\,x^{1/3} + 1.78\,x^{-1/3}.
\end{align}
At $x = 1$ this gives $P(1) = 5.02$~GeV$^4$, in agreement with Eq.~(\ref{eq:pext-eff}). The EOS is again
conformal, $P = \epsilon/3$, as in the main text.

For the conformal EOS at $T = 1$~GeV, the TOV solutions scale with the surface energy density as $M \propto \epsilon_R^{-1/2}$
and $R \propto \epsilon_R^{-1/2}$. Since $\epsilon_R \propto n_R^{4/3}$,
\begin{equation}
\frac{M_{\max}^{\rm eff}}{M_{\max}} =
\frac{R_{\max}^{\rm eff}}{R_{\max}} =
\left(\frac{n_R}{n_R^{\rm eff}}\right)^{2/3}
= \left(\frac{1.793}{1.327}\right)^{2/3} \simeq 1.22.
\end{equation}
Therefore,
\begin{equation}
\,M_{\max}^{\rm eff} \simeq 0.016\,M_\odot, \qquad R_{\max}^{\rm eff} \simeq 107\ \text{m}.
\label{eq:results-eff}
\end{equation}

The shift is a factor of about $1.2$ in both quantities, small compared to the other uncertainties of the model.

\end{document}